\documentclass[12pt,a4paper,BCOR0mm,DIV14,pointlessnumbers,oneside,abstracton]{scrartcl}

\usepackage{mathrsfs}
\usepackage{theorem}
\usepackage[latin1]{inputenc}
\usepackage[center]{caption}
\usepackage{amsfonts}
\usepackage{amstext}
\usepackage{amsmath}
\usepackage{amssymb}
\usepackage{nicefrac}
\usepackage{graphicx}
\usepackage{epic}
\usepackage[komastyle,automark]{scrpage2}
\usepackage{psfrag}
\usepackage{slashed}
\usepackage{feynmp}
\usepackage{cite}
\usepackage{bbm}
\usepackage{pifont}
\usepackage{hyperref}

\DeclareMathOperator{\IM}{Im}

\begin{document}

\flushbottom
\newcommand*\orb[2]{\left[\genfrac{}{}{0pt}{}{#1}{#2}\right]}
\newcommand*{\hwl}{\rule[0mm]{0mm}{4mm}}
\newenvironment*{abstractd}{\renewcommand{\abstractname}{Zusammenfassung}
\begin{abstract}}
{\end{abstract}}

\numberwithin{equation}{section}
\unitlength=1mm 


\titlehead{
\begin{flushright}
TUM-HEP 786/10\\
MPP-2010-170\\
\end{flushright}}

\title{Quark mass hierarchies in heterotic orbifold GUTs}

\author{Rolf Kappl}

\publishers{
\small
Physik-Department T30, Technische Universit\"at M\"unchen,\\
James-Franck-Stra\ss e, 85748 Garching, Germany
\\[5mm]
\small
Max-Planck-Institut f\"ur Physik (Werner-Heisenberg-Institut),\\
F\"ohringer Ring 6,
80805 M\"unchen, Germany
}

\date{}

\maketitle

\begin{abstract}
We discuss how to calculate Yukawa couplings in string-derived
orbifold GUTs. As an application we 
investigate the quark mass hierarchies in these
models. An interplay of different mechanisms derived from
string theory leads to an interesting pattern. We discuss
concrete examples in the context of heterotic orbifold compactifications. 
\end{abstract}

\clearpage



\section{Introduction}

In the standard model (SM) of particle physics 
the structure of Yukawa couplings of
quarks and leptons is completely undetermined. Yukawa couplings 
enter the model as
free parameters. Adjusting these free parameters in an appropriate way, the
theory is extremely successful in describing experimental results.
Unfortunately it cannot explain the differences in the couplings and mixings. It
would be appealing to explain all these input parameters like masses
and mixings from a lower number of free parameters. 

Heterotic string theory is one of the candidates for completing the SM in the
ultraviolet. It was shown in the last years that heterotic orbifold
compactifications can reproduce the particle content of the  SM or its minimal
supersymmetric version (MSSM) \cite{Buchmuller:2005jr,Lebedev:2007hv}.  Yukawa
couplings are in these models no longer free parameters but depend on the size
and the shape of the extra dimensions and the string coupling constant. It is
possible to calculate the Yukawa couplings from first principles and compare
them, as a test of the models, with the experimental data. If such a model is to
describe the real world, it should reproduce the observed quark and lepton
masses, CP phases and mixings. Recently it has been shown that it is possible to
obtain the correct value for the top Yukawa coupling in these models quite
naturally \cite{Hosteins:2009xk}.

Motivated by this success this study is devoted
to a calculation of the Yukawa couplings for the other
particles. It is nearly impossible to calculate the
couplings in string theory exactly due to the complexity of the calculations.
Nevertheless the ratios between the couplings can be computed in a simple way.
We will consider mainly tree-level effects, which means that we will ignore
quantum corrections such as those coming from the renormalization group.

The comparison of the ratios between the different quark masses with the
experimental data is quite exciting. We will show that the so-called benchmark
models 1A and 1B \cite{Lebedev:2007hv}  seem to be excluded due to their flavor
structure. Maybe loop corrections induced through SUSY  breaking
\cite{Buchmuller:1982ye} are able to break the obtained relations in the right
way to get experimentally viable models.

In section \ref{sec:models} we will briefly introduce the mini-landscape models
before we show in section \ref{sec:couplings} how we calculate the ratios
between different couplings. In sections \ref{sec:1A} and \ref{sec:1B} we
elaborate the flavor structure of the models 1A and 1B from
\cite{Lebedev:2007hv} in more detail. Some technical details can be found in
appendix \ref{sec:4point} and \ref{sec:8point}.

\section{The mini-landscape models}
\label{sec:models}

In \cite{Lebedev:2007hv} several string derived models have been
presented which are based on the so-called 
\(\mathbbm{Z}_{6-\text{II}}\) heterotic
orbifold compactification. It was shown that models with 
the exact MSSM spectrum with additional heavy vector-like exotics
exist. These models have appealing phenomenological features like
matter parity, a large top Yukawa coupling \cite{Hosteins:2009xk} 
and the possibility to
include the seesaw mechanism to create neutrino masses 
\cite{Buchmuller:2007zd}. The flavor structure of these models was
also investigated in \cite{Ko:2007dz}. We will focus here on a
specific example and show that the structure of the models is quite
restrictive in a concrete example.

\begin{figure}[ht]
\centering
\includegraphics[width=7cm]{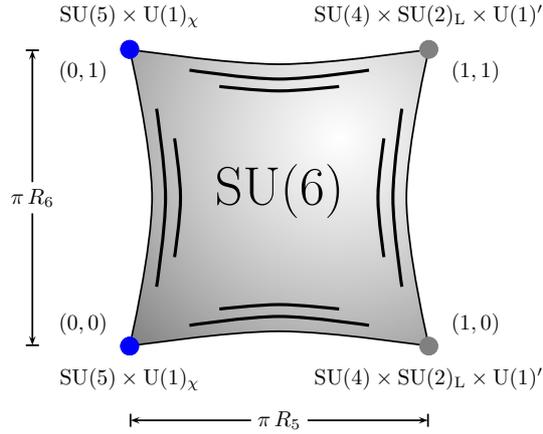}
\caption{A 6D orbifold GUT. The fixed points are labeled with the two
  quantum numbers \((n_2,n_2')\). The first family sits at the fixed point
  labeled with \((0,1)\), the second family at the fixed point
  \((0,0)\). The third family and the Higgs fields live in the bulk.}
\label{fig:SU6}
\end{figure}

We focus on models which allow for a 6D orbifold GUT limit like the ones
considered in \cite{Hosteins:2009xk} or \cite{Buchmuller:2007qf}. Throughout
this analysis we will assume that extra dimensions are compactified in such a
way that this limit is realized. The geometry of the two extra
dimensions is \(\mathbbm{T}^2/\mathbbm{Z}_2\).
The emerging extra dimensional structure
is depicted in figure \ref{fig:SU6}.
The SM families are located at different points in the
extra dimensions (see figure \ref{fig:SU6}). 
These different location properties are not introduced by hand,
but are a consequence of several string selection rules (see for
example \cite{Buchmuller:2006ik}). We will see that the flavor
structure mainly depends on this localization. The fields are further
localized in the additional four extra dimensions needed to have a
consistent string theory. The influence on the flavor
structure of these extra dimensions is subdominant as we will see in the
next section. We can thus completely focus on the 6D GUT structure. A
detailed discussion of the complete geometry can be found for example
in \cite{Buchmuller:2006ik}.

\section{Yukawa couplings in the heterotic mini-landscape}
\label{sec:couplings}

Yukawa couplings in the mini-landscape are not only 3-point functions, but are
induced by higher order terms due to SM singlets. Let us discuss the origin of
the Yukawa couplings in a concrete example. Consider the up type quark
couplings
\begin{equation}
\label{equa:yukaw}
\mathcal{W}_{\text{Yukawa}}\supset 
Y_u^{ij}(R,\tilde{s})q_{i}\overline{u}_j\overline{\phi}_1
\end{equation}
where \(\tilde{s}\) denote standard model singlets, \(R\) represents the
dependence on the extra dimensions and the fields are labeled according to the
notation in \cite{Lebedev:2007hv}. From a bottom-up perspective
\(Y_u^{ij}(R,\tilde{s})\) are free parameters which can be used to generate the
observed quark masses. By contrast, in string theory these couplings are
calculable. 
Matching string scattering amplitudes to
the couplings in the corresponding low-energy field theory results in the 
schematic relation
\begin{equation}
Y_u^{ij}(R,\tilde{s})=
\mathcal{A}_u^{ij}(R)\cdot \text{VEVs}
\end{equation}
where \(\mathcal{A}_u^{ij}(R)\) is the string scattering amplitude. 
The expression ``VEVs'' refers to 
the vacuum expectation value of
SM singlets \(\tilde{s}\) which induce this coupling. In general 
\(Y_u^{ij}(R,\tilde{s})\) is a sum of several contributions (see for example 
equation (\ref{equa:yuk})).
\(\mathcal{A}_u^{ij}(R)\) is the correlation
function of string vertex operators on the string worldsheet 
which can be evaluated with the
help of conformal field theory (CFT). It depends on 
the quantum numbers
of the fields and especially on their origin in the extra dimensions.

Let us specifically consider the so-called benchmark model 1A of
\cite{Lebedev:2007hv}. The Yukawa couplings for the first generation of up
quarks are given by
\begin{equation}
\label{equa:yuk}
\mathcal{A}_u^{11}
=\left<q_1\overline{u}_1\overline{\phi}_1\cdot s^0_4s^0_5s^0_{26}\left(
h_1h_2+s^0_{17}s^0_{18}+s^0_{20}s^0_{22}+s^0_{21}s^0_{22}
+s^0_{20}s^0_{23}+s^0_{21}s^0_{23}
\right)\right>
\end{equation} 
where \(\overline{\phi}_1\) is the up type Higgs field and the additional fields
are SM singlets \((\tilde{s})\). The brackets denote that we have to perform a
CFT calculation to obtain a numerical value. The fields inside this
correlator have to be understood as labels for the corresponding
vertex operators on the string worldsheet.  
Because also the quarks
and the Higgs field participate in this calculation we have included them in the
above formulae. 
To avoid
confusion we want to note that we use the same bracket notation for the VEV
of a field as well as for string correlation functions.
The value of the correlator (\ref{equa:yuk}) together
with the VEVs of the singlets is then the model
prediction for the coefficient \(Y_u^{ij}(R,\tilde{s})\) given in equation
(\ref{equa:yukaw}). The singlet VEVs can be understood as expansion
parameters of the superpotential. 

The calculation can be performed using CFT
techniques. The final result depends only on the
localization of the fields in the extra dimensions.
The reason why only the above outlined couplings are allowed are
several string selection rules specified in detail in
\cite{Buchmuller:2006ik} and 
\cite{Lebedev:2007hv}.

We want to explicitly calculate the above coupling in string theory in
order to explain
or at least  motivate the hierarchy between the quark masses in the SM. The
techniques to calculate such couplings in string theory have been developed long
time ago \cite{Dixon:1986,Hamidi:1986vh}.  However, a technical obstruction to
performing such a computation results from the fact that couplings in
the heterotic
mini-landscape are generated via SM singlets and we have to deal with higher
order couplings. There are calculational tools available also for higher order
couplings \cite{Abel:2003,Atick:1987kd,Bershadsky:1986fv,Choi:2007}, but it is
difficult to obtain numerical quantities from these general results. 
Excited twist operators which appear through oscillator
modes have not been 
considered in the literature for higher order couplings so far. These
couplings occur frequently in the heterotic mini-landscape and it seems that we
have to develop tools to address these couplings as well.

Before getting lost in a lot of technicalities it is worth to look at the
structure of couplings first of all. 
We will see that it is not necessary to develop
new tools, because the coupling structure is quite simple. As we shall see, the
coupling strengths mainly depend on the location and on whether the fields have
oscillator excitations.   
Table \ref{tab:origin} shows the relevant data.
\begin{table}
\centering
\begin{tabular}{ccccccccc}
Particle&Twisted&Oscillator&\(\tilde{N}^{*}_{f_1}\)&\(\tilde{N}^{*}_{f_2}\)&
\(\tilde{N}^{*}_{f_3}\)&\(n_3\)&
\(n_2\)&\(n_2'\)\\
\hline
\(q_1\)&\(\checkmark\)&-&-&-&-&0&0&1\\
\(\overline{u}_1\)&\(\checkmark\)&-&-&-&-&0&0&1\\
\(\overline{\phi}_1\)&-&-&-&-&-&\(*\)&\(*\)&\(*\)\\
\(s^0_4\)&\(\checkmark\)&\(\checkmark\)&-&1&-&0&0&0\\
\(s^0_5\)&\(\checkmark\)&\(\checkmark\)&1&-&-&0&0&0\\
\(s^0_{26}\)&\(\checkmark\)&\(\checkmark\)&-&1&-&0&\(*\)&\(*\)\\
\hline
\(h_1\)&\(\checkmark\)&-&-&-&-&0&\(*\)&\(*\)\\
\(h_2\)&\(\checkmark\)&-&-&-&-&0&\(*\)&\(*\)\\
\(s^0_{17}\)&\(\checkmark\)&-&-&-&-&0&\(*\)&\(*\)\\
\(s^0_{18}\)&\(\checkmark\)&-&-&-&-&0&\(*\)&\(*\)\\
\(s^0_{20}\)&\(\checkmark\)&-&-&-&-&1&\(*\)&\(*\)\\
\(s^0_{21}\)&\(\checkmark\)&-&-&-&-&1&\(*\)&\(*\)\\
\(s^0_{22}\)&\(\checkmark\)&-&-&-&-&2&\(*\)&\(*\)\\
\(s^0_{23}\)&\(\checkmark\)&-&-&-&-&2&\(*\)&\(*\)
\end{tabular}
\caption{We show a set of relevant quantum numbers for some states of the
  model 1A. For the calculation of couplings the geometric origin
and the oscillator numbers are highly important.}
\label{tab:origin}
\end{table}
The notation is the same as in \cite{Lebedev:2007hv}. The six extra
dimensions of the \(\mathbbm{Z}_{6-\text{II}}\) orbifold compactification
factorize in three two dimensional tori
\(\mathbbm{T}^6=\mathbbm{T}^2_{\text{G}_2}\times\mathbbm{T}^2_{\text{SU(3)}}\times
\mathbbm{T}^2_{\text{SO(4)}}\). Each torus is described by a root lattice and an
orbifold point group. As already mentioned we consider the limit in
which the torus with the SO(4) root lattice and the 
\(\mathbbm{Z}_2\) point group is large compared to the other four
extra dimensions.
The quantities 
\(\tilde{N}^{*}_{f_i}\) indicate if a state is from the oscillator sector
in the corresponding torus. Here \(f_1\) refers to the \(G_2\) torus,
whereas \(f_2\) labels the SU(3) torus and \(f_3\) labels the SO(4)
torus. The \(G_2\) and SU(3) torus form four extra dimensions, whereas
the SO(4) torus gives us the two additional extra dimensions depicted
in figure \ref{fig:SU6}. 
No state is an oscillator in the SO(4) torus which simplifies
the calculation as we will see later on. The quantities \(n_2\) and
\(n_2'\) describe the localization in the SO(4) torus as already
discussed and \(n_3\) describes the localization in the SU(3)
torus. A star (\(*\)) means that the state is a bulk field in this torus.
The localization in the \(G_2\) torus is not displayed in the table.

We can see that not all fields are twisted and not every field is an
oscillator state.
As already remarked, computing all involved correlators in string theory
exactly will be painful. A useful strategy to get
phenomenological interesting results has been examined for example in 
\cite{Hosteins:2009xk}. 
Computing a quantity exactly is often quite difficult, but
the relation to another observable is sometimes very easy to
obtain. We will use a similar strategy here. We do not calculate the
Yukawa couplings exactly, but look at the ratios between them. As we
can see by looking at the correlators in string theory, 
it is irrelevant where a
term comes from, for example from a singlet or a quark. The 
only important point is where the fields are located. 
The reason for this behavior is of course that in heterotic string theory the
different particles are just excitations of a closed string in ten
dimensions. Because we are considering the interactions between
closed strings we already incorporate all essential features. An
additional advantage of this approach is that we do not have to take
care about the K\"ahler potential because it is universal for the
different couplings (see also \cite{Ibanez:1992hc}).

To be
more concrete, let us take a look at another coupling. We find
\begin{equation}
\mathcal{A}_u^{12}=
\left<q_1\overline{u}_2\overline{\phi}_1\cdot s^0_4s^0_{11}s^0_{26}\left(
h_1h_2+s^0_{17}s^0_{18}+s^0_{20}s^0_{22}+s^0_{21}s^0_{22}
+s^0_{20}s^0_{23}+s^0_{21}s^0_{23}
\right)\right>.
\end{equation}
The structure is the same as for the \(Y_u^{11}\) coupling except for the
fact that \(s_{5}^0\) was interchanged by \(s_{11}^0\). These singlets
have the same quantum numbers (as can be seen in the appendix of 
\cite{Lebedev:2007hv}), 
but live at different fixed points (see also table
\ref{tab:origin}). This is not an accident but a consequence of a
\(D_4\) symmetry \cite{Ko:2007dz}.
The only difference between \(\overline{u}_1\) and \(\overline{u}_2\)
is, that they also live at different fixed points. This means, that the
interchange in the singlets just compensates the different location of
the second family. We get
\begin{equation}
Y_u^{12}=
\frac{\left<s_{11}^0\right>}{\left<s_5^0\right>}Y_u^{11}.
\end{equation}
With  \(\left<\tilde{s}\right>\) we denote the vacuum
expectation value (VEV) of a field \(\tilde{s}\).
We obtain similar relations like that, which only depend on the
VEVs of some singlet fields and not on some difficult string theory
calculation. Using these relations we can determine the structure of
the Yukawa couplings in dependence of the yet unknown quantity
\(Y_u^{11}\).

\section{The flavor structure of model 1A}
\label{sec:1A}

\subsection{The up quark sector}

We will consider the up quark sector in the so-called benchmark model 1A in
the heterotic mini-landscape in this section. In principle, 
this analysis can easily
extended to all models in the mini-landscape. In \cite{Lebedev:2007hv}
the couplings in the up quark sector were given by
\begin{equation}
Y_u=\begin{pmatrix}
\tilde{s}^5&\tilde{s}^5&\tilde{s}^5\\
\tilde{s}^5&\tilde{s}^5&\tilde{s}^6\\
\tilde{s}^6&\tilde{s}^6&1
\end{pmatrix}.
\end{equation}
We have already looked in more detail at the couplings \(Y_u^{11}\)
and \(Y_u^{12}\). The entry \(Y_u^{33}\) was set to 1 because it
already appears at the renormalizable level. In \cite{Hosteins:2009xk}
it was shown that this coupling comes from the gauge coupling in the
higher dimensional gauge theory and is slightly suppressed due to
localization effects to fit in the right value. For our purposes we
can just assume that the coupling to the top quark is of the right
size and set it in the following to \(y_t\).

We have collected the behavior of all necessary fields in table
\ref{tab:origin} and table \ref{tab:fields}.
\begin{table}
\centering
\begin{tabular}{ccccccccc}
Particle&Twisted&Oscillator&\(\tilde{N}^{*}_{f_1}\)&\(\tilde{N}^{*}_{f_2}\)&
\(\tilde{N}^{*}_{f_3}\)&\(n_3\)&
\(n_2\)&\(n_2'\)\\
\hline
\(q_2\)&\(\checkmark\)&-&-&-&-&0&0&0\\
\(\overline{u}_2\)&\(\checkmark\)&-&-&-&-&0&0&0\\
\(q_3\)&-&-&-&-&-&\(*\)&\(*\)&\(*\)\\
\(\overline{u}_3\)&-&-&-&-&-&\(*\)&\(*\)&\(*\)\\
\(s^0_6\)&\(\checkmark\)&\(\checkmark\)&1&-&-&0&0&0\\
\(s^0_9\)&\(\checkmark\)&\(\checkmark\)&-&1&-&0&0&1\\
\(s^0_{11}\)&\(\checkmark\)&\(\checkmark\)&1&-&-&0&0&1\\
\end{tabular}
\caption{More information on the participating fields in the up quark sector.}
\label{tab:fields}
\end{table}
We start with a closer look at the coupling \(Y_u^{11}\). As can be
seen from the quantum numbers in table \ref{tab:origin} the fields are
located at different fixed points in the SO(4) torus depicted in
figure \ref{fig:SU6}. This different
location will be encoded in the quantity \(f(R_6)\)
\begin{equation}
Y_u^{11}=f(R_6)A(\tilde{s}).
\end{equation} 
We will look at this quantity in more
detail later (see also appendix \ref{sec:4point}). We will see that
the quark mass hierarchies are mainly controlled by \(f(R_6)\). The
quantity \(A(\tilde{s})\) is no longer depending on the localization
of the fields in the SO(4) torus, but still on the localization in the
other extra dimensions and on the singlet VEVs. 
The localization in the additional four extra dimensions 
will be irrelevant for our
discussion and is therefore suppressed. The reason for that is that 
\(A(\tilde{s})\) appears in every entry of \(Y_u\) and can therefore
not lead to a difference between the quark couplings.

Looking at the other couplings in detail we get
\begin{equation}
Y_u^{21}=Y_u^{12},\qquad Y_u^{22}=A(\tilde{s})
\end{equation}
where we have taken 
the different localization of the second family in \(Y_u^{22}\) into
account.
We have
the same fields as for the \(Y_u^{11}\) coupling despite the two
quarks which are located at a different fixed point. For the
\(Y_u^{22}\) coupling all
fields are located at the same fixed point in the SO(4) torus and we
get no contribution from the localization in this torus. The coupling
is completely insensitive to the volume of this torus.

When including also the third family things get more complicated.
The third family is not localized and has its 
origin in the untwisted sector (see table 
\ref{tab:fields}). In order to obtain couplings to localized states, extra
singlet VEVs are required, such that the couplings appear at higher order.
We obtain for example
\begin{equation}
\begin{split}
\mathcal{A}_u^{13}&=\left<q_1\overline{u}_3\overline{\phi}_1\cdot s^0_4s^0_6s^0_{11}s^0_{26}\left(
h_1h_2+s^0_{17}s^0_{18}+s^0_{20}s^0_{22}+s^0_{21}s^0_{22}
+s^0_{20}s^0_{23}+s^0_{21}s^0_{23}\right)\right>
\\
&\quad +
\left<q_1\overline{u}_3\overline{\phi}_1\cdot s_{13}^0
\left(s^0_4s^0_{26}\right)^2\right>.
\end{split}
\end{equation}
Compared to \(Y_{11}\) the singlet \(s_6^0\) is here playing the role
of \(s_5^0\) whereas \(s_{11}^0\) is playing the role of
\(\overline{u}_1\). Because \(\overline{u}_3\) is from the untwisted
sector it is irrelevant for our considerations. There is
an additional term at order five in the singlets which we cannot easily
relate to \(Y_u^{11}\). We obtain
\begin{equation}
Y_u^{13}=\frac{\left<s_{6}^0\right>\left<s_{11}^0\right>}
{\left<s_5^0\right>}Y_u^{11}+B(R_6,\tilde{s}),
\qquad Y_u^{31}=\left<s_{9}^0\right>Y_u^{11}
\end{equation}
where \(B(R_6,\tilde{s})\) stands for an unknown contribution
appearing at order five in the singlets. Furthermore we have
\begin{equation}
Y_u^{23}=\left<s_{6}^0\right>A(\tilde{s}),\qquad Y_u^{32}=
\frac{\left<s_{9}^0\right>\left<s_{11}^0\right>}
{\left<s_5^0\right>}Y_u^{11}.
\end{equation}
The last term
occurs already at the renormalizable level and we set it to
\(Y_u^{33}=y_t\) as already discussed. 
Putting everything together we get the matrix
\begin{equation}
Y_{u}=\begin{pmatrix}
f(R_6)A(\tilde{s})&\frac{\left<s_{11}^0\right>}{\left<s_5^0\right>}
f(R_6)A(\tilde{s})&
\frac{\left<s_{6}^0\right>\left<s_{11}^0\right>}
{\left<s_5^0\right>}f(R_6)A(\tilde{s})+B(R_6,\tilde{s})\\
\frac{\left<s_{11}^0\right>}{\left<s_5^0\right>}f(R_6)A(\tilde{s})&
A(\tilde{s})&\left<s_{6}^0\right>A(\tilde{s})\\
\left<s_{9}^0\right>f(R_6)A(\tilde{s})
&\frac{\left<s_{9}^0\right>\left<s_{11}^0\right>}
{\left<s_5^0\right>}f(R_6)A(\tilde{s})&y_t
\end{pmatrix}.
\end{equation}
This matrix contains several parameters. The singlet VEVs are, in principle,
calculable as they are determined by F- and D-flatness constraints. However, in
practice it is quite difficult to compute them. In what follows, we will see
that one can nevertheless gain non-trivial information since we can understand
the behavior of \(f(R_6)\) from a string theory
calculation.
Another important information is that the diagonal entries of \(Y_u\) are
completely independent of the VEVs. Let us now discuss the quantities \(f(R_6)\)
and \(B(R_6,\tilde{s})\) in somewhat more detail.


The expression \(f(R_6)\) reflects the different localization 
of the first family in
the SO(4) torus. It thus depends only on 
the moduli dependent part of the amplitude
in this torus. We have four twisted fields and no oscillator. Looking at the
location of the fields we can see that for \(Y_u^{22}\) all fields live at the
same fixed point in this torus. We conclude that there is no suppression due to
instanton effects (see for example \cite{Dixon:1986}).  For \(Y_u^{11}\) the
situation changes. We get differently localized fields and thus an instanton
suppression. As a first result we can assume that \(Y_u^{22}\gg
Y_u^{11}\) and therefore \(f(R_6)\ll 1\). 

Whenever two fields from
different fixed points interact the coupling is suppressed by the
distance between the fields. If only fields from the same orbifold
fixed point or from the bulk interact the coupling is unsuppressed.   

A detailed look at a four twist calculation (as for example performed
in
\cite{Dixon:1986}) tells us that the only
difference between \(Y_u^{11}\) and \(Y_u^{22}\) is in the classical part
of the amplitude. This is the instanton part which goes roughly like
\(e^{-R_6^2}\) where \(\pi R_6\) is the distance between the two fixed points
labeled with \(n_2'\) in the SO(4) torus (of course this distance is
encoded in the \(T\) modulus). 
We
get the following suppression (see appendix \ref{sec:4point} for a detailed 
calculation)
\begin{equation}
\label{equa:T}
f(R_6)= e^{-\frac{\pi R_6^2}{4}}.
\end{equation}
The quantity \(B(R_6,\tilde{s})\) is quite difficult to discuss. It is 
a coupling
involving six oscillators. It is also the only coupling
depending on \(s^0_{13}\). \(B(R_6,\tilde{s})\) appears only in one
entry of the Yukawa matrix and thus it
affects the result only marginaly.

Using this ingredients we can try to obtain quantitative
results. The running of the couplings is neglected as it is
beyond the accuracy of our approach (see also \cite{Hosteins:2009xk}). 
As a first test, the eigenvalues of \(Y_u\) should give us the
right ratio of the different up type quark masses. 
We start with fixed VEVs, all of the same
size \(\left<s^0_{i}\right>=0.1\). 
As usual we can change to 
mass eigenstates to relate the couplings in \(Y_u\) to the quark
masses. The diagonal entries of the rotated matrix have to be proportional to
the quark masses. The experimental values are given in table
\ref{tab:exup}.
\begin{table}
\centering
\begin{tabular}{cc}
Quark&Mass\\
\hline
up quark (\(\overline{u}_1\))&\(1.5-3.3 \text{ MeV}\)\\
charm quark (\(\overline{u}_2\))
&\(1.27\begin{pmatrix}+0.07\\ -0.11\end{pmatrix} 
\text{ GeV}\)\\
top quark (\(\overline{u}_3\))&\(171.3 \pm 1.1 \pm 1.2 \text{ GeV}\)
\end{tabular}
\caption{Up quark masses from \cite{Amsler:2008zzb}.}
\label{tab:exup}
\end{table}
We get ratios between the quark masses like 500
between the first two generations and 100 between the second and the
third generation
\begin{equation}
\frac{m_u}{m_c}\stackrel{!}{\approx} \frac{1}{500},\qquad \frac{m_c}{m_t}
\stackrel{!}{\approx} 
\frac{1}{100}.
\end{equation}

The impact of the VEVs and \(B(R_6,\tilde{s})\) is negligible and we
ignore it. 
We therefore set \(B(R_6,\tilde{s})\) to zero. 

The dependence on the remaining quantities is approximately given by
\begin{equation}
\frac{m_u}{m_c}\approx f(R_6),\qquad \frac{m_c}{m_t}\approx A(\tilde{s}).
\end{equation}
If we choose 
\(f(R_6)=\frac{1}{500}\) we can get the right behavior with 
\(A(\tilde{s})=\frac{1}{100}\). The off-diagonal terms are now quite small
and we obtain the eigenvalues \(m\approx (0.00002, 0.01, 1)\). Given
the experimental values (displayed in table \ref{tab:exup}) these
values are of the right size.

The small value of \(f(R_6)\) can be explained by the instanton
suppression in the SO(4) torus. Using our result from equation 
(\ref{equa:T}) we obtain \(R_6\approx 2.8\). As also remarked in 
\cite{Hosteins:2009xk} we expect to relate the distance \(R_5\) between
the two fixed points labeled by \(n_2\) to the GUT scale, namely
\(M_{\text{GUT}}=\frac{1}{2R_5}\). As a starting point we use \(R_5=50\),
which gives us \(M_{\text{GUT}}=\frac{1}{100}\) in Planck units. With
these values we
find an anisotropy of a factor of \(\frac{R_5}{R_6}=18\) 
to get the right value for \(R_6\). Given \cite{Hosteins:2009xk}
such a value seems to be excluded. Nevertheless it is
very encouraging that we get a value in the correct region. As outlined
in section \ref{sec:dquark} the model suffers from more serious problems,
therefore we do not address this point here in more detail.
Another uncertainty is the value for
\(A(\tilde{s})\) which may or may not be realistic.

We conclude that even if we have a lot of unknowns, 
we can use the strong correlations between the couplings due to
the different selection rules to make predictions. In contrast to the
general analysis of 
\cite{Ko:2007dz} we find that the models are quite predictive.
It is
appealing that we have to consider in principle only a 6D GUT with
differently localized families to explain the hierarchy for the up quark
masses. We have obtained a very intuitive geometrical explanation and no
miraculous string features at work.  
Taking the result of \cite{Hosteins:2009xk} into account the scenario
is very constraint. We have one
free parameter, namely \(R_6\), to explain the ratio between the
gauge coupling and the top Yukawa coupling and simultaneously the
ratio between the up and the charm quark mass.

We proceed by looking at the remaining quark masses in
the down quark sector.

\subsection{The down quark sector}
\label{sec:dquark}

After the success of relating the quark masses in the up quark sector
to the geometry of the extra dimensions, we will also try to explain the
hierarchy for the down quark masses on the same
footing. As we will see one cannot fit the up quark and the down quark
masses simultaneously. The experimental data are
displayed in table \ref{tab:exdown}.
\begin{table}
\centering
\begin{tabular}{cc}
Quark&Mass\\
\hline
down quark (\(\overline{d}_1\))&\(3.5-6 \text{ MeV}\)\\
strange quark (\(\overline{d}_2\))&\(105\begin{pmatrix}+25\\ -35\end{pmatrix} 
\text{ MeV}\)\\
bottom quark (\(\overline{d}_3-\overline{d}_4
\))&\(4.2\begin{pmatrix}+0.17\\ -0.07\end{pmatrix} 
\text{ GeV}\)
\end{tabular}
\caption{Down quark masses from \cite{Amsler:2008zzb}. The labels for
  the quark masses are
  the same as in \cite{Lebedev:2007hv}.}
\label{tab:exdown}
\end{table}
We find a mass ratio of 19 between the 
down and the strange quark mass \cite{Leutwyler:1996qg} and
a ratio around 40 between the strange and the bottom quark mass
\begin{equation}
\frac{m_d}{m_s}\stackrel{!}{\approx} \frac{1}{19},\qquad \frac{m_s}{m_b}
\stackrel{!}{\approx} 
\frac{1}{40}.
\end{equation}
We take into account couplings up to order nine
in standard model singlets which correspond to 12-point
couplings. Because we consider also couplings at higher order in singlets our
starting point is slightly different from \cite{Lebedev:2007hv} and we
start with
\begin{equation}
Y_d=\begin{pmatrix}
\tilde{s}^9&\tilde{s}^5&0\\
\tilde{s}^5&\tilde{s}^9&0\\
0&\tilde{s}^6&\tilde{s}^8
\end{pmatrix}
\end{equation}
which is a more accurate starting point.
Repeating the steps of the discussion in the up quark sector we end up with 
a matrix
like
\begin{equation}
Y_d=\begin{pmatrix}
g_1(R_6)C(\tilde{s})&f(R_6)D(\tilde{s})&0\\
f(R_6)D(\tilde{s})&g_2(R_6)C(\tilde{s})&0\\
0&X(R_6,\tilde{s})&U(\tilde{s})
\end{pmatrix}.
\end{equation} 
We have a lot of 
parameters, 
from a naive perspective it should be
easily possible to fit the right values for all quark
masses.

For the coupling between \(q_3\) and
the massless combination \(\overline{d}_3-\overline{d}_4\) we find a
coupling at order eight which is unsuppressed in the SO(4) torus because
we find a coupling where all fields are localized at the same fixed
point. Looking at the eigenvalues of the matrix we can see that the
coupling \(U(\tilde{s})\) 
exactly gives us one eigenvalue. As outlined above
the coupling is not related to \(R_6\) or \(R_5\) and we can treat it as
a free parameter to fit the bottom quark mass. The situation is
similarly as for the top quark coupling, the third generation decouples
from the first two generations.

For the other two quark masses the situation is
more complicated. 
We find that \(X(R_6,\tilde{s})\) plays no role for the eigenvalues of
\(Y_d\) and is thus irrelevant for our discussion. The off-diagonal
couplings \(f(R_6)D(\tilde{s})\) arise 
already at order five in standard model
singlets and are suppressed in the SO(4) torus. We find the
same factor as in the up quark sector, namely \(f(R_6)\) for the
suppression. 
\(f(R_6)\) is the same as in the
up quark sector, we have again four twisted
fields in the SO(4) torus. Again two fields are
localized at the fixed point labeled by \(n_2'=0\), whereas two of
them are localized at the fixed point \(n_2'=1\). To be concrete, we obtain
\begin{equation}
\mathcal{A}_{d}^{12}=
\left<q_1\overline{d}_2\phi_1\cdot s_6^0s_9^0s_{26}^0
\left(
h_1h_2+s^0_{17}s^0_{18}+s^0_{20}s^0_{22}+s^0_{21}s^0_{22}
+s^0_{20}s^0_{23}+s^0_{21}s^0_{23}
\right)\right>
+\mathcal{O}(\tilde{s}^6)
\end{equation}
and thus
\begin{equation}
Y_{d}^{12}=f(R_6)D(\tilde{s})=\frac{1}
{\tan\beta}
\frac{\left<s_6^0\right>\left<s_9^0\right>}{\left<s_4^0\right>
\left<s_5^0\right>}Y_u^{11}+\mathcal{O}(\tilde{s}^6).
\end{equation}
Inserting already the Higgs VEVs we can relate quantities from the up quark
sector to the down quark sector. We have seen that
\(Y_u^{11}=f(R_6)A(\tilde{s})\) together with \(\tan\beta\)
gives the up quark mass. We thus
can see that \(Y_d^{12}\) is of the right order to give us the down
quark mass because the singlet VEVs are supposed to be 
of the same order of magnitude (see for example
\cite{Kappl:2008ie}). Of course, in detail the situation is more
constrained, 
because \(\tan\beta=1\) is excluded by experiments. An allowed value
of about \(\tan\beta > 2\) \cite{Degrassi:2002fi} 
yields a too small down quark
mass if the VEVs are of the same size. Nevertheless, this can be
overcome due to an according VEV assignment. If this can be realistic or
not is an open question.

We further note that we get contributions from
higher order terms. These corrections do not affect the
statement \(Y_d^{12}=Y_d^{21}\) which is correct also up to
order nine in the standard model singlets. The reason for this behavior is the
\(D_4\) symmetry \cite{Ko:2007dz}.

The diagonal
couplings \(g_1(R_6)C(\tilde{s})\) and \(g_2(R_6)C(\tilde{s})\) 
arise at order nine in standard model singlets. Their calculation is
challenging.
We have (for the
quantum numbers of the additional fields see \cite{Lebedev:2007hv})
\begin{equation}
Y_d^{11}=g_1(R_6)C(\tilde{s}),\qquad Y_d^{22}=\frac{g_2(R_6)}{g_1(R_6)}Y_d^{11}
\end{equation}
with
\begin{equation}
\begin{split}
\mathcal{A}_d^{11}&=\left<q_1\overline{d}_1\phi_1\cdot s_4^0\left(s_{13}^0
\right)^2s_{26}^0s_{30}^0
\big(
\chi_1\chi_2h_2h_4+\chi_3\chi_4h_2h_4+
\chi_1\chi_4h_2h_6+\chi_2\chi_3h_2h_6\right.\\
&\quad \left.+
h_2h_3(h_4)^2+h_2h_3(h_6)^2+h_2h_4h_5h_6
\right)\Big>.
\end{split}
\end{equation}
We find a
similar behavior as in the up quark sector. The couplings for the
first and second generation are not the same, but differ due to the
different localization of the second family. 
We introduce the factors \(g_1(R_6)\) and \(g_2(R_6)\) which encode
the localization behavior. 
We
find for these couplings 
eight fields twisted in the SO(4) torus. There are always fields
localized at different fixed points. We conclude that despite the fact
that the coupling is already suppressed by a high power of singlet
VEVs, the coupling is further suppressed due to the localization in
the SO(4) torus (see also appendix \ref{sec:8point}). 
Compared to the off-diagonal terms, these couplings
are therefore rather small. We have to introduce two
different factors because both couplings are suppressed. In the up
quark case only the \(Y_u^{11}\) coupling was suppressed.

%
We want to look at the eigenvalues to compare them to the experimental
quark masses like in the up quark case. \(X(R_6,\tilde{s})\)
is irrelevant for the eigenvalues due to the zeros in the
matrix \(Y_d\).
Because the off-diagonal entries are identical and compared to the
diagonal couplings quite large we get two almost degenerate
eigenvalues. This is in contrast to data. To be more concrete we
can estimate the difference between the couplings. If we assume rather
large VEVs \(\left<\tilde{s}\right>=0.3\) we get a suppression of 
\(\left<\tilde{s}\right>^4\approx \frac{1}{125}\) between the diagonal and the 
off-diagonal couplings because the diagonal couplings arise at order
nine and the off-diagonal couplings at order five in standard model singlets 
\begin{equation}
C(\tilde{s})\approx \frac{1}{125}D(\tilde{s}).
\end{equation}
We should point out that for smaller VEVs this ratio is even
larger. 
If we use the results of appendix \ref{sec:8point} we obtain
\begin{equation}
Y_d\approx\begin{pmatrix}
\frac{1}{125}f(R_6)D(\tilde{s})&f(R_6)D(\tilde{s})&0\\
f(R_6)D(\tilde{s})&\frac{1}{125}f(R_6)D(\tilde{s})&0\\
0&X(R_6,\tilde{s})&U(\tilde{s})
\end{pmatrix}.
\end{equation} 
We can tune \(D(\tilde{s})\) to obtain the right ratio between the
bottom and the strange quark mass regardless of \(R_6\), 
but it is impossible to get a big ratio
between the down and the strange quark mass. The quark mass ratios are
\begin{equation}
\frac{m_d}{m_s}\approx \frac{62}{63},
\qquad \frac{m_s}{m_b}\approx \frac{126}{125}\frac{f(R_6)D(\tilde{s})}
{U(\tilde{s})}.
\end{equation}
The mass eigenvalues for \(m_d\) and \(m_s\) are
almost degenerate for all possible values of the free parameters.

We conclude that it is
impossible to obtain a quark mass hierarchy as expected from
experimental data with our setup. 

It was first proposed in \cite{Buchmuller:1982ye} to generate Yukawa
couplings via loop effects through SUSY breaking soft terms. This
can be a mechanism to cure the problem in the down quark sector. This
mechanism leads also naturally to flavor changing neutral currents (FCNCs)
which have to be compatible with experimental bounds (see for example
\cite{Banks:1987iu} in the string theory context). If it is possible
to fulfill these bounds in this class of models remains an open task
for future work.
Maybe there are also other mechanism
which can give rise to essential corrections to the down quark masses
in the right way. 

\section{A comment on model 1B}
\label{sec:1B}

In \cite{Lebedev:2007hv} another vacuum configuration called
model 1B was considered. This model has an unbroken \(D_4\) family
symmetry for the first two generations. The difference to model 1A
is that the symmetry is not broken by the VEV assignment.
The consequence is that all
couplings for the first two generations are equal. As we have seen in
the last section for the model 1A, such a behavior is in conflict
with observations. It
seems to be impossible for such configurations to get a big ratio between the
masses as required by experiment without radiative quark mass
generation already mentioned in the last section. 
We conclude that these
configurations are rather unrealistic by construction.

\section{Summary}
\label{sec:summary}

We have developed tools to study the quark mass pattern in heterotic
orbifold compactifications. Specifically we 
have studied the flavor structure in two examples of the
mini-landscape, namely the so-called benchmark models 1A and 1B. We have
seen that without any difficult string theory calculation we can
obtain several interesting results because the couplings enjoy
relations. 
We have
linked the quark masses to the extra dimensions like already in 
\cite{Hosteins:2009xk} for the top quark coupling. We have also found
some relations between the up quark and the down quark sector
depending on \(\tan \beta\). Our study shows that it is possible to
obtain simple relations for the Yukawa couplings even in complex
string-derived setups. 

For the concrete models studied in our analysis we found
that it is impossible to get the 
right masses in the down quark sector without 
including large radiative contributions through SUSY breaking.
We do not know if this feature is
universal for all models in the mini-landscape. What we can state is
that it seems to be quite easy to disfavor models from the
mini-landscape. Because we have not succeeded in obtaining the right quark
mass pattern, we have not looked at the mixing, namely the CKM matrix
or the mass pattern in the lepton sector. With the techniques outlined
in this paper this is in principle also
possible.

In \cite{Buchmuller:2008uq} and \cite{Petersen:2009ip} powerful 
tools have been developed to study
the discrete symmetries between the string selection rules. Maybe it
is possible to find better string theory models with these techniques.

\section*{Acknowledgments}

We would like to thank Michael Ratz for discussion and helpful
comments on the manuscript. This research was supported by the DFG
cluster of excellence Origin and Structure of the Universe, and the
SFB-Transregio 27 "Neutrinos and Beyond" by Deutsche
Forschungsgemeinschaft (DFG).

\appendix

\section{$\boldsymbol{N}$-point couplings for four $
\boldsymbol{\mathbbm{Z}_2}$ twists}
\label{sec:4point}

To obtain the quantity \(f(R_6)\) we need to calculate a
\(N\)-point coupling with four \(\mathbbm{Z}_2\) twists. The calculation
of a 4-point coupling with four \(\mathbbm{Z}_2\) twists has been
calculated long time ago \cite{Dixon:1986,Hamidi:1986vh}. Let us
review here the result. There are two different contributions, one
from the quantum part and one from the classical part. The first is
insensitive to the localization of the fields, whereas the latter one
is sensitive. Thus the quantity \(f(R_6)\) is only influenced by the
classical part. The result for the classical part is \cite{Dixon:1986}
\begin{equation}
Z_{\text{cl}}(x)=\sum e^{-S_{\text{cl}}(x)}
=\sum_{n_0,n_1\in\mathbbm{Z}}e^{-\frac{\pi R_6^2}{\IM\tau(x)}
\left|n_1+n_0\tau(x)+\frac{1}{2}(\epsilon_1+\epsilon_0\tau(x))\right|^2} 
\end{equation}   
where we have
\begin{equation}
\tau(x)=\frac{i\,{}_2F_1\left(\frac{1}{2},\frac{1}{2},1,1-x\right)}
{{}_2F_1\left(\frac{1}{2},\frac{1}{2},1,x\right)}=
\frac{i\,K(1-x)}{K(x)}
\end{equation}
and \(\epsilon_i\) labels the fixed point of the twist fields. With
\(K(x)\) we mean the complete elliptic integral of the first kind. 
\(x\) can be
interpreted as the position of one of the vertex operators on the
sphere. It cannot be fixed by \(\text{PSL}(2,\mathbbm{C})\)
invariance and has to be weighted by an integral in the end. That means we have
one free complex modulus of the Riemann surface describing 
the string interaction. We further have \(\epsilon_i=0\) for the fixed point 
\((n_2,n_2')=(0,0)\) and \(\epsilon_i=1\) for the fixed point 
\((n_2,n_2')=(0,1)\).

If all fields live at the same fixed point \(\epsilon_0=\epsilon_1=0\)
we get an unsuppressed amplitude because the exponential breaks
down. If the fields are localized at different fixed points, for
example \(\epsilon_0=0\) and \(\epsilon_1=1\) we get a suppression
going with the distance squared \(R_6^2\). The terms for \(n_i\neq 0\)
are further suppressed and come from states winding also around the
torus. We can neglect their contribution for simplicity and obtain
\begin{equation}
\label{equa:clas}
Z_{\text{cl}}(x)\approx \begin{cases}
1 & \text{all fields live at the same fixed point}\\
e^{-\frac{\pi R_6^2}{4\IM\tau(x)}} & \text{fields live at different fixed points} 
\end{cases}.
\end{equation}
This is the exact classical contribution in the SO(4) torus, because
all other fields are untwisted in the SO(4) torus. 
We do not know the contribution in the other extra dimensions, namely
the SU(3) and the \(\text{G}_2\) torus in
detail. We know that the additional contribution is
depending on \(x\). Because we have to integrate over \(x\) in the
end, we have to approximate the result.

\begin{figure}
\centering
\psfrag{rx}{\small{\(\mathsf{\textsf{Re}(x)}\)}}
\psfrag{ix}{\small{\(\mathsf{\textsf{Im}(x)}\)}}
\psfrag{S}{\small{\(\mathsf{S_{\textsf{cl}}(x)}\)}}
\includegraphics[width=8cm]{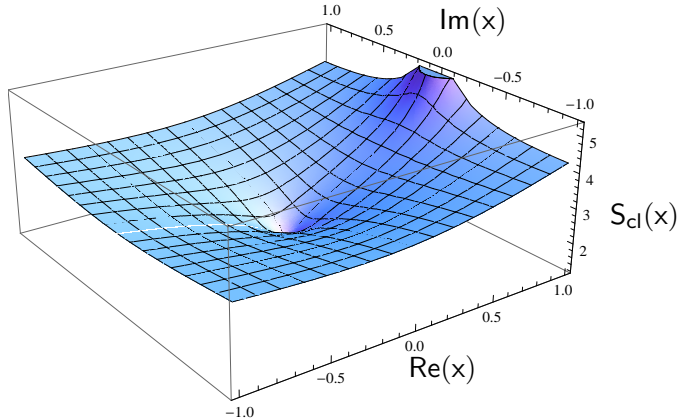}
\caption{The function \(S_{\text{cl}}(x)\) for the values
  \(\epsilon_0=0\), \(\epsilon_1=1\) and \(R_6=2\).}
\label{fig:clas}
\end{figure}

Let us focus now on the \(N\)-point coupling in more
detail. We have a \(N\)-point coupling and therefore \(N\) vertex operator
positions on the worldsheet. We can fix three of them by 
\(\text{PSL}(2,\mathbbm{C})\) invariance. Over the remaining \(N-3\)
positions we should integrate. At least one of these vertex operators
is twisted in the SO(4) torus 
and we called his position \(x\) (in fact it is also possible that all
vertex operators in the SO(4) torus are undetermined 
by \(\text{PSL}(2,\mathbbm{C})\) invariance, but we will not consider
this case here). The other vertex operators are
untwisted in the SO(4) torus and we do not know their contribution
exactly, but they depend on all \(N-3\) variables.
Their contribution is the same, regardless of their localization 
in the SO(4) torus. The complete contribution coming from this
localization enters via equation (\ref{equa:clas}). If we call the
unknown part from the SU(3) and \(\text{G}_2\) torus and the
contributions not affected by the localization in the SO(4) torus 
\(U(x,x_2,\ldots x_{N-3})\) we get
\begin{equation}
\intop d^2x \intop d^2x_2 \ldots \intop d^2x_{N-3} Z_{\text{cl}}(x) 
U(x,x_2,\ldots x_{N-3})
\end{equation}
for the complete amplitude.
We can approximate the function \(Z_{\text{cl}}(x)\) by
setting \(x=\frac{1}{2}\) which is motivated
by the
shape of the function (see figure \ref{fig:clas}). 
We obtain \(\tau(x)=i\) for this choice. We get
\begin{equation}
\intop d^2x \intop d^2x_2 \ldots \intop d^2x_{N-3} 
U(x,x_2,\ldots x_{N-3})
\end{equation} 
if all fields live at the same fixed point and the coupling is
unsuppressed. In the case where the fields are localized at different
fixed points we get
\begin{equation}
e^{-\frac{\pi R_6^2}{4}}\intop d^2x \intop d^2x_2 \ldots \intop d^2x_{N-3} 
U(x,x_2,\ldots x_{N-3}).
\end{equation} 
We recognize that
the integrals are now the same. We conclude that the required
ratio between an suppressed and an unsuppressed coupling is
\begin{equation}
f(R_6)=\frac{e^{-\frac{\pi R_6^2}{4}}\intop d^2x \intop d^2x_2 \ldots 
\intop d^2x_{N-3} 
U(x,x_2,\ldots x_{N-3})}{\intop d^2x \intop d^2x_2 \ldots \intop d^2x_{N-3} 
U(x,x_2,\ldots x_{N-3})}=e^{-\frac{\pi R_6^2}{4}}
\end{equation}
in a first approximation. The uncertainty of this approach is of
course substantial, 
because the function \(U(x,x_2,\ldots x_{N-3})\) is unknown.

\section{$\boldsymbol{N}$-point couplings for eight $
\boldsymbol{\mathbbm{Z}_2}$ twists}
\label{sec:8point}

The functions \(g_1(R_6)\) and \(g_2(R_6)\) 
can be determined by looking at \(N\)-point
couplings with eight $\mathbbm{Z}_2$ twists. The computation of this
amplitude is more complicated than the one reviewed in appendix
\ref{sec:4point}. General $\mathbbm{Z}_2$ twist amplitudes
have been considered in several papers 
\cite{Atick:1987kd,AlvarezGaume:1986es,
Dijkgraaf:1987vp,Knizhnik:1987xp,Miki:1987mp,Zamolodchikov:1987ae},
also 
because they
are of interest in the context of spin
operators. The beautiful mathematics of hyperelliptic
Riemann surfaces needed for these computations can be found for example in
\cite{farkas:riemann}. The factors \(g_1(R_6)\) and \(g_2(R_6)\) 
are not only given by
the classical part of the amplitude, which is now depending on at
least five free moduli of the Riemann surface \(x_1,\ldots x_5\). The
reason for this is that different fields appear for different couplings 
and the quantum part is depending
on all of these fields. Because our approach contains already a big
uncertainty we assume 
\begin{equation}
Z_{\text{qu}}=\mathcal{O}(1)
\end{equation}
and neglect the quantum part for the following discussion. This can be
justified due to the discussion in \cite{Abe:2009dr} in the context of
magnetized brane models and will be discussed in more detail for
heterotic orbifolds elsewhere.

The
classical part can be written as
\begin{equation}
Z_{\text{cl}}=\sum e^{-S_{\text{cl}}(x_1,\ldots ,x_5)}.
\end{equation}
The dependence on \(R_6\) is
the same as for \(f(R_6)\).
We find
\begin{equation}
g_i(R_6)=e^{-\alpha_i R_6^2}
\end{equation}
and thus again an exponential suppression of the order of
\(f(R_6)\). The constants \(\alpha_i > 0\) are depending on the period
matrix of the Riemann surface and on the integral
approximation. The coupling \(g_1(R_6)\) is dominated by couplings
where six twist fields are localized at one fixed point and two twist
fields at the other fixed point. 
We thus conclude that \(g_1(R_6)\approx f(R_6)\) (see for example 
\cite{Atick:1987kd}). For the coupling \(g_2(R_6)\) the dominant
contribution originates from the same coupling as for the \(g_1(R_6)\) term.
There is an additional contribution
 where four twist fields sit at
one fixed point and four twist fields at the other 
fixed point. Nevertheless, 
this contribution is stronger suppressed 
\cite{Atick:1987kd}. We thus conclude
that \(g_2(R_6)\approx g_1(R_6)\). There is only a mild difference
between \(g_2(R_6)\) and \(g_1(R_6)\) which we will neglect.
For the numerical discussion of couplings we assume
\begin{equation}
g_1(R_6)\approx g_2(R_6)\approx f(R_6).
\end{equation}
The error of this approximation is larger as for the calculation
of \(f(R_6)\) discussed in appendix \ref{sec:4point}.

\bibliographystyle{ArXiv}
\bibliography{mini}

\end{document}